\documentclass[prd, nofootinbib, superscriptaddress]{revtex4-2}
\usepackage[utf8]{inputenc}
\usepackage[T1]{fontenc}
\usepackage[english]{babel}

\usepackage{amsmath, amssymb, amsfonts, mathtools}
\usepackage{slashed, braket, bm} 
\usepackage{float, graphicx}
\usepackage{xcolor, url, hyperref}

\definecolor{ref_blue}{rgb}{0, 0, 0.9}
\hypersetup{colorlinks = true,
	linkcolor = ref_blue,
	citecolor = ref_blue,		
	filecolor = ref_blue,		
	urlcolor = ref_blue}	

\newcommand\nt{\notag}
\renewcommand\d{{\rm d}}
\newcommand\e{{\rm e}}

\newcommand{\varpar}{{\mkern3mu\vphantom{\perp}\vrule depth 0pt\mkern2mu\vrule depth 0pt\mkern3mu}}
\newcommand\qTfv{q_{\scriptscriptstyle T}}
\newcommand\pTfv{p_{\scriptscriptstyle T}}
\newcommand\bTfv{b_{\scriptscriptstyle T}}
\newcommand\qT{{\bm q}_{\scriptscriptstyle T}}
\newcommand\pT{{\bm p}_{\scriptscriptstyle T}}
\newcommand\kT{{\bm k}_{\scriptscriptstyle T}}
\newcommand\bT{{\bm b}_{\scriptscriptstyle T}}
\newcommand\xB{x_{\scriptscriptstyle B}}
\newcommand\muH{\mu_{\scriptscriptstyle H}}
\begin{document}
\title{Transverse momentum dependent shape function for \texorpdfstring{$J/\psi$}{\textit{J/ψ}} production in SIDIS}

\author{Daniël Boer}
\email{d.boer@rug.nl}
\affiliation{Van Swinderen Institute for Particle Physics and Gravity, University of Groningen, Nijenborgh 4, 9747 AG Groningen, The Netherlands}

\author{Jelle Bor}
\email{j.bor@rug.nl}
\affiliation{Van Swinderen Institute for Particle Physics and Gravity, University of Groningen, Nijenborgh 4, 9747 AG Groningen, The Netherlands}
\affiliation{IJCLab, CNRS, Université Paris-Saclay, 91405 Orsay, France}

\author{Luca Maxia}
\email{l.maxia@rug.nl}
\affiliation{Van Swinderen Institute for Particle Physics and Gravity, University of Groningen, Nijenborgh 4, 9747 AG Groningen, The Netherlands}

\author{Cristian Pisano}
\email{cristian.pisano@unica.it}
\affiliation{Dipartimento di Fisica, Università di Cagliari, Cittadella Universitaria, I-09042 Monserrato (CA), Italy}
\affiliation{INFN, Sezione di Cagliari, Cittadella Universitaria, I-09042 Monserrato (CA), Italy}

\author{Feng Yuan}
\email{fyuan@lbl.gov}
\affiliation{Nuclear Science Division, Lawrence Berkeley National Laboratory, Berkeley, CA 94720, USA}

\begin{abstract}
It has been shown previously that the transverse momentum dependent (TMD) factorization of heavy quarkonium production requires a TMD shape function. Its perturbative tail can be extracted by matching the cross sections valid at low and high transverse momenta. In this article we compare the order-$\alpha_s$ TMD expressions with the order-$\alpha_s^2$ collinear ones for $J/\psi$ production in semi-inclusive deep inelastic scattering (SIDIS), employing nonrelativistic QCD in both cases.
In contrast to previous studies, we find that the small transverse momentum limit of the collinear expressions contains discontinuities.
We demonstrate how to properly deal with them and include their finite contributions to the TMD shape functions. Moreover, we show that soft gluon emission from the low transverse momentum Born diagrams provide the same leading order TMD shape functions as required for the matching. Their revised perturbative tails have a less divergent {behavior} as compared to the TMD fragmentation functions of light hadrons. 
Finally, we investigate the universality of TMD shape functions in heavy quarkonium production, identify the need for process dependent factorization and discuss the phenomenological implications.
\end{abstract}

\date{\today}
\maketitle

\section{Introduction}
In recent years, heavy quarkonium production in various inclusive processes has attracted great interest as a way to probe the {transverse momentum dependent (TMD)} gluon distributions~\cite{Godbole:2012bx, Boer:2012bt, Godbole:2013bca, Dunnen:2014eta, Mukherjee:2015smo, Mukherjee:2016cjw, Mukherjee:2016qxa, Rajesh:2018qks, Scarpa:2019fol, DAlesio:2019:AzAsEIC, Kishore:2021vsm}. In this paper, we focus on $J/\psi$ production in semi-inclusive deep inelastic scattering (SIDIS), 
\begin{equation}
    e\, (\ell) + p\, (P) \to e^\prime\, (\ell^\prime) + \gamma^* \, (q) + p\, (P) \to e^\prime\, (\ell^\prime) + J/\psi \, (P_\psi) + X\ ,
\label{eq: SIDIS process}
\end{equation}
where the particle momenta are given between brackets and the virtual photon momentum is given by $q = \ell - \ell'$. 
The $J/\psi$ mass $M_\psi^2 = P_\psi^2$ and the photon virtuality $Q^2 = - q^2>0$ are considered hard scales in the process, {i.e.}\ they are considered much larger than the nonperturbative QCD scale $\Lambda_{\text{QCD}}$, although most results will also be valid for photoproduction ($Q^2=0$). The electron and proton masses will be neglected w.r.t.\ $M_\psi$ and $Q$ whenever possible. The virtual photon transverse momentum is denoted by $\qT$ and can be directly related to the $J/\psi$ transverse momentum $\bm P_{\psi \perp}$.
{The distinct subscripts used for the transverse momentum components, specifically “$T$” and “$\perp$”, serve to emphasize the different frames in which they are measured.
In particular, we consider $\qT$ when both the target proton and the $J/\psi$ have no transverse components and $\bm P_{\psi \perp}$ when the photon and the proton have only longitudinal components.}

\begin{figure}[t]
    \centering
    \includegraphics[width=0.8\linewidth, keepaspectratio]{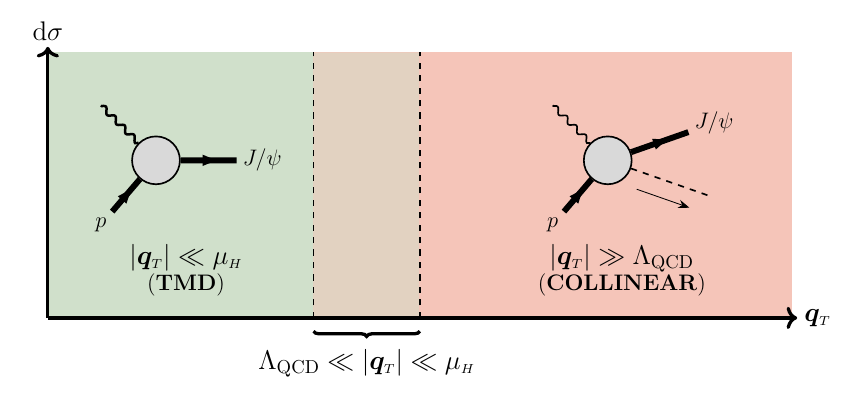}
    \caption{Schematical overview of matching in $ep\to e' J/\psi + X$ to obtain the leading order shape function.}
    \label{FigMatch}
\end{figure}

Depending on the value of $|\qT|$, we can identify two different transverse momentum regions, see Fig.~\ref{FigMatch}. The high transverse momentum (HTM) region is given by the condition $|\qT| \gg \Lambda_{\rm QCD}$, while the low transverse momentum (LTM) region corresponds to $|\qT| \ll \muH$. Here $\muH = f(Q, M_\psi)$ {with $f(Q, M_\psi) \gtrsim M_\psi$} generically denotes the hard scale of the process.
The cross section can be evaluated within the two transverse momentum regions by adopting the proper factorization that enables to separate the short-distance from the long-distance contributions.
The collinear factorization is applicable at HTM, while the TMD factorization is expected to be valid at LTM~\cite{Bacchetta:2018:GluonTMDsEIC}, for which the cross section is sensitive to {TMD} quantities.
{In addition, we can identify an intermediate transverse momentum (ITM) region, namely $\Lambda_{\rm QCD} \ll |\qT| \ll \muH$, where both factorizations are valid. Since our attention will be mostly directed towards this overlapping region, where $|\qT|$ (or equivalently $|\bm P_{\psi\perp}|$) becomes small compared to the hard scale, we will neglect any transverse momentum dependence in $f(Q, M_\psi)$.}

To describe $J/\psi$ hadronization we employ nonrelativistic~QCD (NRQCD) \cite{Bodwin:1994:NRQCD}, in which the heavy-quark pair forms a Fock state, specified by $n={}^{2S+1} L_J^{[c]}$: $S$ denotes the spin, $L$ the {orbital} angular momentum, $J$ the total angular momentum and $c$ the color {state} of the {pair}. Note that the pair can couple either as a color-singlet (CS), {with $c = 1$,} or as a color-octet (CO) state, {with $c = 8$}.  
The (low-energy) transition from this general state to the $J/\psi$ is encoded in the nonperturbative Long-Distance Matrix Elements (LDMEs) that are distinct for each quarkonium Fock state. 
States with different quantum numbers $n$ do not interfere as the cross section is proportional to a direct sum of LDMEs, up to a required precision in the expansion {w.r.t.~$v$, which corresponds to the (non-relativistic) relative velocity of the heavy quark-antiquark pair in the quarkonium rest frame.}
In the following we will truncate the expansion up to the relative order $v^4$, including the {${}^3 S_1^{[1]}$} CS state  and the {${}^1 S_0^{[8]}$, ${}^3 S_1^{[8]}$, ${}^3 P_J^{[8]}$} CO states. {Note that, in the following we will not consider the interference among $P$-wave states since it is not necessary in the evaluation of the unpolarized differential cross section. However, we have taken them into account in our brief digression on the production of polarized $J/\psi$ mesons in SIDIS (see Sec.~\ref{sec: Conclusions}).}

In Refs.~\cite{Echevarria:2019:TMDShF, Fleming:2019:TMDShF} it was found that the TMD factorized expressions have to take into account final state smearing effects that are encoded in the TMD shape function (TMDShF). 
This nonperturbative hadronic quantity describes the transition from the heavy quark pair to a bound quarkonium state, which not only contains the formation of the bound state in terms of an LDME, but also the transverse momentum effects that arise from the soft-gluon radiation.

In Refs.~\cite{Boer:2020:epJpsiMatching, DAlesio:2021:epJpsiMatchingPol} the matching procedure in SIDIS has been investigated, according to which the TMD and collinear expressions are compared in the {ITM region.} It was found that the introduction of TMDShFs solves the mismatch between the collinear and TMD expressions, by resumming $|\qT|$ divergences in the Sudakov factor. However, this term is in contradiction with other studies, as it has been demonstrated that no double logarithms in the nonperturbative Sudakov factor associated to heavy quark production are present for $pp \to (J/\psi\ {\rm or}\ \Upsilon) +X$ \cite{Sun:2012:ppJpsiResum} and for open heavy-quark pair production, both in $ep$ \cite{Zhu:2013:epOpenQuark} and $pp$ \cite{Zhu:2012:OpenTopHadronCollider} collisions. 
The absence of the double logarithms in $J/\psi$ production can also be seen in Ref.~\cite{Echevarria:2022:talk@Tansversity}.
Due to this discrepancy, universality was assumed in \cite{Bor:2022:EvolTMDShF} using the explicit result of \cite{Sun:2012:ppJpsiResum}; however, as we will see, this only holds for photoproduction, not electroproduction.

We found that the discrepancy in the $ep$ matching study arises from the presence of discontinuities in the structure functions that appear in the small-$\qT$ limit of collinear factorized expressions. These structure functions contain a Dirac delta function for which a small-$\qT$ approximation is applied. The approximation employed in Refs.~\cite{Boer:2020:epJpsiMatching, DAlesio:2021:epJpsiMatchingPol}, which is an extension of a well-known expression \cite{Meng:1995:SIDISsmallqT} to the heavy quarkonium case, would be valid when multiplied by a continuous function, but that turns out to be invalid in the present case of discontinuous hard scattering factors. 
In this article we show how to properly treat these expressions to resolve this discrepancy.

In addition, we extend our analysis to single quarkonium production in $pp$ collisions, where the hard scale is a function of the quarkonium mass only: $\muH = f(M_\psi)$ {with $f(M_\psi) \sim M_\psi$}.
This allows us to test the connection of TMDShFs obtained in different cases, {i.e.}~to study their universal properties.
Even if we expect that the LDMEs are process independent in the collinear description, the same is not necessarily true for the TMDShFs. Indeed in the latter process dependences may arise due to the transverse momentum exchange with other colored objects.

The paper is organized as follows.
In Sec.~\ref{sec: matching procedure} we revise the matching procedure. In particular, in Sec.~\ref{sec: matching High -> Int} we discuss the pole structure of the collinear cross section in the small transverse momentum limit in detail, while the new TMDShF results for SIDIS are presented in Sec.~\ref{sec: matching Low -> Int}.
In Sec.~\ref{sec: Universality} we address the aforementioned process dependence, comparing the TMDShFs in SIDIS and in $pp$ collisions.
Conclusions are given in Sec.~\ref{sec: Conclusions}, together with a summary of our findings.
{In addition, there are two appendices at the end of this paper. In Appendix~\ref{app: poles} we present a more complete derivation of our method to include the pole structure contributions in our results. In Appendix~\ref{app: eikonal method} we derive the soft gluon emission from the Born amplitude obtained through the eikonal approximation.}

\section{The matching procedure}
\label{sec: matching procedure}
The SIDIS reaction in Eq.~\eqref{eq: SIDIS process} is described by the conventional kinematical SIDIS variables
\begin{equation}
    \xB = \frac{Q^2}{2 P \cdot q}\ , \quad y = \frac{P \cdot q}{P \cdot \ell}\ , \quad z = \frac{P \cdot P_\psi}{P \cdot q}\ .
\end{equation}
We  consider a frame where the virtual photon has no transverse momentum component, and we identify two light-cone directions $n_+$ and $n_-$, for which $n_+ \cdot n_- = 1$. With these, the Sudakov decomposition of the relevant momenta can be written as
\begin{equation}
\begin{aligned}
    P^\mu & = n_+^\mu\ , \\
    q^\mu & = - \xB n_+^\mu + \frac{Q^2}{2 \xB} n_-^\mu\ , \\
    P_\psi^\mu & = \frac{ \xB M_{\psi \perp}^2}{z Q^2}n_+^\mu + \frac{z Q^2}{2\xB} n_-^\mu + {P}_{\psi \perp}^\mu\ , 
\end{aligned}
\end{equation}
where ${P}_{\psi \perp}^2 = - {\bm P}_{\psi \perp}^2$ is the squared $J/\psi$ transverse momentum (w.r.t.\ the photon and proton), while $M_{\psi \perp} = \sqrt{M_\psi^2 + {\bm P}_{\psi \perp}^2}$ is the $J/\psi$ transverse mass.

In particular, we will consider the fully unpolarized differential cross section ${\d \sigma/(\d \xB\, \d y\, \d z\, \d \qT^2\, \d \phi_\psi})$,
where $\phi_\psi$ is the $J/\psi$ azimuthal angle measured w.r.t.\ the lepton plane. Moreover, we replaced the transverse momentum of the $J/\psi$ with that of the photon $\qT$ 
(evaluated w.r.t.\ the hadrons); this replacement is achieved via
\begin{equation}
    |\qT| = \frac{1}{z} |{\bm P}_{\psi \perp}|\ .
\end{equation}
The differential cross section can be parameterised in the HTM region as follows \cite{Boer:2020:epJpsiMatching}
\begin{align}
    \frac{\d \sigma}{\d \xB\, \d y\, \d z\, \d \qT^2\, \d \phi_\psi} & = \frac{\alpha}{y Q^2} \bigg\{ \Big[1 + (1 - y)^2 \Big] \, F_{UU, \perp} + 4\, (1 - y)\, F_{UU, \varpar} \nt \\ 
    & \phantom{=} + 2\, (2 - y) \sqrt{1 - y}\, \cos \phi_\psi\, F_{UU}^{\cos\phi_\psi} + 4\, (1 - y)\, \cos 2 \phi_\psi\, F_{UU}^{\cos 2 \phi_\psi} \bigg\}\ ,
\label{eq: SIDIS diff cross section}
\end{align}
where the first two subscripts of the structure functions $F$ refer to the polarization of the initial (unpolarized) proton and electron. 
The last subscript in $F_{UU,{\cal P}}$ with ${\cal P} = \perp,\, \varpar$ refers to the virtual photon polarization (transverse or longitudinal), while for $F_{UU}^{\Phi}$ with $\Phi = \cos \phi_\psi,\, \cos 2\phi_\psi$ the superscript refers to the angular term that accompanies it.
Henceforth, we will refer to the aforementioned hard scattering structure functions via the general notation $F_{UU,{\cal P}}^\Phi$.
On the other hand, the same differential cross section evaluated in the LTM region is given by
\begin{align}
    \frac{\d \sigma}{\d \xB\, \d y\, \d z\, \d \qT^2\, \d \phi_\psi} & = \frac{\alpha}{y Q^2} \bigg\{ \Big[1 + (1 - y)^2 \Big] \, {\cal F}_{UU, \perp} + 4\, (1 - y)\, {\cal F}_{UU, \varpar} + 4\, (1 - y)\, \cos 2 \phi_\psi\, {\cal F}_{UU}^{\cos 2 \phi_\psi} \bigg\}\ ,
\label{eq: SIDIS diff cross section TMD}
\end{align}
{where the structure function ${\cal F}_{UU}^{\cos \phi_\psi}$, being subleading power/twist, has not been included.}
Note the difference in the structure functions: $F_{UU,{\cal P}}^\Phi$ are evaluated in collinear factorization, while the calligraphic ${\cal F}_{UU,{\cal P}}^\Phi$ are calculated within transverse momentum factorization.

\subsection{From high to intermediate transverse momentum}
\label{sec: matching High -> Int}
In this section we provide a systematic method to investigate the small-$\qT$ limit of SIDIS observables at HTM ({${|\qT| \gg \Lambda_{\rm QCD}}$}). Adopting the parton model, the production of a $J/\psi$ possessing a high transverse momentum component is possible at the lowest order in $\alpha_s$ via
\begin{equation}
    \gamma^* \, (q) + a \, (p_a) \to c\bar c[n]\, (P_\psi) + a^\prime\, (p_a^\prime)\ ,
\label{eq: partonic interaction HTM}
\end{equation}
where $a$ can be either a quark, antiquark, or gluon. In this kinematical regime we adopt collinear factorization, for which
\begin{equation}
    p_a \approx \xi P\ .
\end{equation}

The perturbative amplitude squared $|{\cal M}|^2$ for the hadronic process in Eq.~\eqref{eq: SIDIS process} is obtained by contracting the leptonic tensor $L^{\mu\nu}$ with the amplitude $H_{\mu}^{(a)\, [n]}$, describing the partonic process in Eq.~\eqref{eq: partonic interaction HTM}, and its conjugate $H_{\nu}^{(a)\, [n]*}$. 
In particular, the lepton tensor can be written as follows:
\begin{equation}
    L^{\mu\nu} = \frac{Q^2}{y^2} \bigg\{ \Big[1 + (1 - y)^2 \Big] \, \epsilon^{\mu\nu}_{\perp} + 4\,(1 - y)\,  \left( \epsilon^{\mu\nu}_{\varpar} + \epsilon^{\mu\nu}_{\cos 2 \phi_\psi} \right) + 2\, (2 - y) \sqrt{1 - y} \, \epsilon^{\mu\nu}_{\cos\phi_\psi}  \bigg\} \ ,
\end{equation}
where we introduced the tensors
\begin{equation}
    \epsilon^{\mu\nu}_{\perp} = - g_\perp^{\mu\nu} \ , \quad
    \epsilon^{\mu\nu}_{\varpar} = \epsilon_L^\mu\, \epsilon_L^\nu \ , \quad
    {\epsilon^{\mu\nu}_{\cos \phi_\psi} =
    \left( \epsilon_L^\mu \hat \ell_\perp^\nu + \hat \ell_\perp^\mu \epsilon_L^\nu \right)\ , \quad
    \epsilon^{\mu\nu}_{\cos 2 \phi_\psi} = 
    \left( \hat \ell_\perp^\mu \hat \ell_\perp^\nu + \frac{1}{2} g_\perp^{\mu\nu}\right)}\ .
\label{eq: lepton tensors def}
\end{equation}
Moreover, $g_\perp^{\mu\nu}$ is the transverse projector
\begin{equation}
    g_\perp^{\mu\nu} \equiv g^{\mu\nu} - \frac{1}{P \cdot q} \left( P^\mu q^\nu + q^\mu P^\nu \right) - \frac{Q^2}{\left( P \cdot q\right)^2} P^\mu P^\nu\ ,
\end{equation}
while $\epsilon_L^\mu(q)$ is the longitudinal polarization vector
\begin{equation}
    \epsilon_L^\mu(q) = \frac{1}{Q} \left( q^\mu + \frac{Q^2}{P\cdot q} P^\mu\right)\ ,
\end{equation}
and $\hat \ell^\mu_\perp$ is the unit vector along the transverse component of $\ell$, w.r.t.\ the photon-proton axis.
Henceforth, we refer to one of the tensors in Eq.~\eqref{eq: lepton tensors def} via the general notation $\epsilon^{\mu\nu}_{{\cal P};\Phi}$, where ${\cal P} = \perp,\, \varpar$ and $\Phi = \cos \phi_\psi,\, \cos 2\phi_\psi$.
Employing this, the structure functions introduced in Eq.~\eqref{eq: SIDIS diff cross section} can be evaluated via
\begin{align}
    F_{UU,{\cal P}}^\Phi 
    & = \frac{1}{4\left(4\pi \right)^3} z \sum_{n} \sum_{a} \int_{\xB}^{\hat x_{\rm max}} \frac{\d \hat x}{\hat x} \int_z^1 \frac{\d \hat z}{\hat z} \, \frac{1}{Q^2}\, {f_1^a}(\xi; \mu^2)\, \epsilon^{\mu\nu}_{{\cal P};\Phi} \, H_{\mu}^{(a)\, [n]}H_{\nu}^{(a)\, [n]*}\, \langle {\cal O}[n] \rangle \nt \\
    & \phantom{=} \times \delta \bigg(\frac{(1 - \hat x)(1 - \hat z)}{\hat x \hat z} - \frac{1 - \hat z}{\hat z^2}\, \frac{M_\psi^2}{Q^2} - \frac{\qT^2}{Q^2} \bigg)\, \delta(z - \hat z) \ ,
\label{eq: structure function partonic definition}
\end{align}
where the sum $n$ runs over the dominant LDMEs $\langle {\cal O}[n] \rangle$ and $a$ runs over the parton types. Furthermore, we introduced the partonic scaling variables
\begin{equation}
    \hat x = \frac{Q^2}{2 p_a \cdot q} = \frac{\xB}{\xi}\ , \quad \hat z = \frac{p_a \cdot P_\psi}{p_a \cdot q} = z\ ,
\label{eq: partonic variables definition}
\end{equation}
together with
\begin{equation}
    \hat x_{\rm max} = \frac{Q^2}{M_\psi^2 + Q^2}\ .
\end{equation}
In Ref.~\cite{Boer:2020:epJpsiMatching} the Dirac delta present in Eq.~\eqref{eq: structure function partonic definition} was expanded at small-$\qT$ as follows (see its Appendix~B for the derivation)
\begin{align}
    \delta \bigg(\frac{(1 - \hat x)(1 - \hat z)}{\hat x \hat z} - \frac{1 - \hat z}{\hat z^2}\, \frac{M_\psi^2}{Q^2} - \frac{\qT^2}{Q^2} \bigg)
    & \sim \hat x_{\rm max} \bigg[ \log \frac{M_\psi^2 + Q^2}{\qT^2}\,\delta(1 - \hat x')\, \delta(1 - \hat z) \nt\\
    & \phantom{=} + \frac{\hat x'}{(1 - \hat x')_+}\, \delta(1 - \hat z) + \frac{M_\psi^2 + Q^2}{M_\psi^2/\hat z + Q^2}\frac{\hat z}{(1 - \hat z)_+}\, \delta(1 - \hat x') \bigg]\ ,
\label{eq: naive delta expansion}
\end{align}
where 
\begin{equation}
    \hat x' = \frac{\hat x}{\hat x_{\rm max}}\ .
\label{eq: x hat prime definition}
\end{equation}
Note that on the right-hand side of Eq.~\eqref{eq: naive delta expansion} the coefficient in front of the double delta logarithmically diverges with $\qT$.
However, as we have found this is not sufficient to obtain the correct behavior of structure functions in the ITM region, which is restored by adding a 
constant term to the double-delta coefficient.\footnote{
Note that this new term has the same divergence order as the other terms on right-hand side of Eq.~\eqref{eq: naive delta expansion}, namely the coefficients of $\delta(1 - \hat x')$ and $\delta(1 - \hat z)$.
}
The need to include this subdominant term can also be understood from the following argument. 
The Dirac-delta expansion in Eq.~\eqref{eq: naive delta expansion} was obtained in Ref.~\cite{Boer:2020:epJpsiMatching} by applying the full Dirac delta to two continuous test functions. However, the structure functions defined in Eq.~\eqref{eq: structure function partonic definition} contain discontinuities {that come from the soft gluon radiation associated with the CO final state in the NRQCD calculations (see Appendix~\ref{app: eikonal method}). These contributions} are  made explicit via the decomposition into poles through a Laurent expansion, 
namely\footnote{
Before performing the expansion, we suggest applying once the following relation, obtained from the Dirac delta: $$\frac{1}{1 -\hat z} = \frac{(\hat z - \hat x')\, M_\psi^2 + \hat z\,(1 - \hat x')\, Q^2}{\hat x'\,  \hat z^2\, \qT^2}\ .$$ 
}
\begin{equation}
    \frac{1}{\left( 4 \pi \right)^3} \frac{1}{Q^2} \epsilon_{{\cal P};\, \Phi}^{\mu\nu}\, H_{\mu}^{(a)\, [n]} H_{\nu}^{(a)\, [n]*} \equiv 
    {\cal H}_{{\cal P};\, \Phi}^{(a)\, [n]}(\hat x^\prime, \hat z) = 
    {\cal H}_{{\cal P};\, \Phi}^{(a)\, [n];\, {(0)}}(\hat x^\prime, \hat z) 
        + \sum_{k=1}^\infty \left( \frac{1 - \hat z}{1 - \hat x^\prime} \right)^k {\cal H}_{{\cal P};\, \Phi}^{(a)\, [n];\, (k)} (\hat z)\ ,
\label{eq: structure functions Laurent expansion}
\end{equation}
where ${\cal H}_{{\cal P};\, \Phi}^{(a)\, [n];\, {(0)}}$ and all ${\cal H}_{{\cal P};\, \Phi}^{(a)\, [n];\, {(k)}}$ are finite.
{Despite the different notation, the amplitude squared on the left-hand side of Eq.~\eqref{eq: structure functions Laurent expansion} is in agreement with Refs.~\cite{Kniehl:2001tk, Sun:2017nly}. Note that to get the pole structure on the right-hand side of Eq.~\eqref{eq: structure functions Laurent expansion} we are explicitly writing the amplitude squared in terms of $\hat x'$ and $\hat z$ (Eqs.~\eqref{eq: partonic variables definition} and~\eqref{eq: x hat prime definition}).
{We have found that for $J/\psi$ production in SIDIS the poles are present only for the gluon-initiated process $\gamma^* g$ with the expansion running up to $k = 2$. 
Instead, the quark-initiated processes $\gamma^* q$ are fully described by the $k=0$ finite term.
Moreover, up to the precision considered in this work, the poles contribute only to the structure functions $F_{UU,{\cal P}}$ introduced in Eq.~\eqref{eq: SIDIS diff cross section}.}

These poles are under control when the amplitude squared is evaluated at high-$\qT$ values, as the transverse momentum forces the phase space to deviate from $\hat z = 1$ and $\hat x' = 1$.
Solely when we consider the small-$\qT$ limit they have a significant impact.}
The Dirac-delta expansion in Eq.~\eqref{eq: naive delta expansion} is applicable only to the first term (${\cal H}_{{\cal P},\, \Phi}^{(a)\, [n];\,(0)}$), while all the others require a different approach.
{In particular}, we can split the differential cross section in Eq.~\eqref{eq: SIDIS diff cross section} into three parts in the HTM region, namely
\begin{equation}
    \frac{\d \sigma}{\d \xB\, \d y\, \d z\, \d \qT^2\, \d \phi_\psi} \equiv \d \sigma_{\scriptscriptstyle A} + \d \sigma_{\scriptscriptstyle B} + \d \sigma_{\scriptscriptstyle C}\ ,
\end{equation}
with
\begin{equation}
\begin{aligned}
    \d \sigma_{\scriptscriptstyle A} 
    & = \frac{\alpha}{4 y Q^2}\, z \sum_n \int_0^1 \d \hat x' \int_0^1 \d \hat z \, {f_1^a} \bigg(\frac{\xB}{\hat x_{\rm max}\, \hat x'}; \mu^2\bigg) \, \hat{z} 
    \ \delta\big( G(\hat x', \hat z) \big) \, \delta(z - \hat z) \\ & \phantom{=} \times 
        \bigg\{ 
            \Big[1 + (1 - y)^2 \Big] \, {\cal H}_{\perp}^{(a)\, [n];\, (0)}(\hat x', \hat z) + 4\,(1 - y)\, {\cal H}_{\varpar}^{(a)\, [n];\, (0)}(\hat x', \hat z)\\ 
        & \phantom{=} 
            + 2\, (2 - y) \sqrt{1 - y}\, \cos \phi_\psi\, {\cal H}^{(a)\, [n];\, (0)}_{\cos\phi_\psi}(\hat x', \hat z) 
            + 4\, (1 - y)\, \cos 2 \phi_\psi\, {\cal H}^{(a)\, [n];\, (0)}_{\cos 2 \phi_\psi}(\hat x', \hat z) \bigg\} \,
            \langle {\cal O}[n] \rangle\ 
            \ , \\
    \d \sigma_{\scriptscriptstyle B} 
    & = \frac{\alpha}{4 y Q^2} {z} \sum_n \int_{0}^{1} {\d \hat x'} \int_0^1 {\d \hat z}\, {f_1^g} \bigg(\frac{\xB}{\hat x_{\rm max}\, \hat x'}; \mu^2\bigg)\, \hat{z} \ \delta\big( G(\hat x', \hat z) \big) \, \delta(z - \hat z)  \\ 
        & \phantom{=} \times \left( \frac{1 - \hat z}{1 - \hat x^\prime} \right) \left[ \Big(1 + (1 - y)^2 \Big) \, {\cal H}_{\perp}^{(g)\, [n];\, (1)}(\hat z) + 4\, (1 - y)\, {\cal H}_{\varpar}^{(g)\, [n];\, (1)}(\hat z) \right]
            \langle {\cal O}[n] \rangle\ , \\
    \d \sigma_{\scriptscriptstyle C}
    & = \frac{\alpha}{4 y Q^2} {z} \sum_n \int_{0}^{1} {\d \hat x'} \int_0^1 {\d \hat z}\, {f_1^g} \bigg(\frac{\xB}{\hat x_{\rm max}\, \hat x'}; \mu^2\bigg) \, \hat{z} \ \delta\big( G(\hat x', \hat z) \big) \, \delta(z - \hat z) \\ 
        & \phantom{=} \times \left( \frac{1 - \hat z}{1 - \hat x^\prime} \right)^2 \left[ \Big(1 + (1 - y)^2 \Big) \, {\cal H}_{\perp}^{(g)\, [n];\, (2)}(\hat z) + 4\, (1 - y)\, {\cal H}_{\varpar}^{(g)\, [n];\, (2)}(\hat z) \right] 
            \langle {\cal O}[n] \rangle \ ,
\label{eq: differential cross section separation}
\end{aligned}
\end{equation}
where the function $G(\hat x', \hat z)$ is given by
\begin{equation}
    G(\hat x', \hat z) = \hat z\, ( 1 - \hat z) (1 - \hat x') - \frac{M_\psi^2}{Q^2}(1 - \hat z)(\hat x' - \hat z) - \frac{\qT^2}{Q^2}\hat x'\, \hat z^2\, .
\label{eq: full delta G}
\end{equation}
{The difference in the lower integration limit for both $\hat x'$ and $\hat z$ between Eqs.~\eqref{eq: structure function partonic definition} and~\eqref{eq: differential cross section separation} has been introduced for the convenience of the calculation. This modification is possible since the added integration range does not contribute to the final result (see Appendix~\ref{app: poles} of this work and Appendix~B of \cite{Boer:2020:epJpsiMatching}).}
As mentioned, one can apply directly the delta expansion in Eq.~\eqref{eq: naive delta expansion} to evaluate the small-$\qT$ behavior of $\d \sigma_A$.
Instead, the expansions of $\d \sigma_B$ and $\d \sigma_C$ are obtained by considering the integral w.r.t.\ $\d \hat x$ and $\d \hat z$ of those terms that are truly indeterminate in the limit $\qT\to0$, with the indeterminacy solved by the presence of the full Dirac delta (see Eq.~\eqref{eq: full delta G}). 

Therefore, it is legitimate to approximate 
${\cal H}_{\cal P}^{[n];\, (k)} (\hat z) \to {\cal H}_{\cal P}^{[n];\, (k)}(1)$
which gives
\begin{equation}
\begin{aligned}
    {\cal H}_{\cal P}^{(g)\, [n];\, (1)}(1) & = - 2\, \frac{M_\psi^2}{M_\psi^2 + Q^2} {\cal H}_{\cal P}^{(g)\, [n];\, (0)}(1, 1)\ , \\
    {\cal H}_{\cal P}^{(g)\, [n];\, (2)}(1) & = \left(\frac{M_\psi^2}{M_\psi^2 + Q^2} \right)^2 {\cal H}_{\cal P}^{(g)\, [n];\, (0)}(1, 1)\ .
\end{aligned}
\end{equation}
Hence from Eq.~\eqref{eq: differential cross section separation} we obtain 
\begin{align}
    \d \sigma_{\scriptscriptstyle B} & \approx \frac{\alpha}{4 y Q^2}  {f_1^g} (x; \mu^2) \left( - \hat x_{\rm max}\, \log \frac{M_\psi^2}{\qT^2} \right) \nt \\ & \phantom{\approx} \times \sum_n \left[ \Big(1 + (1 - y)^2 \Big) \, {\cal H}_{\perp}^{(g)\, [n];\, (0)}(1,1) + 4\, (1 - y)\, {\cal H}_{\varpar}^{(g)\, [n];\, (0)}(1,1) \right] \langle {\cal O}[n] \rangle \, \delta(1 - z) \, 
\label{eq: sigma B approximation}
\end{align}
and
\begin{align}
    \d \sigma_{\scriptscriptstyle C} & \approx \frac{\alpha}{4 y Q^2} {f_1^g} (x; \mu^2) \left[ \frac{\hat x_{\rm max}}{2} \left( \log \frac{M_\psi^2}{\qT^2} - 1 \right) \right] \nt \\ & \phantom{\approx} \times \sum_n \left[ \Big(1 + (1 - y)^2 \Big) \, {\cal H}_{\perp}^{(g)\, [n];\, (0)}(1,1) + 4\, (1 - y)\, {\cal H}_{\varpar}^{(g)\, [n];\, (0)}(1,1) \right] \langle {\cal O}[n] \rangle \, \delta(1 - z) \ ,
\label{eq: sigma C approximation}
\end{align}
where $x \equiv \xB/\hat x_{\rm max}$. 
More details on the previous results can be found in Appendix~\ref{app: poles}.
Since the small-$\qT$ limit of these quantities is proportional to ${\cal H}^{[n];\, (0)}(1,1)$, we can effectively add these terms to the double delta coefficient, obtaining that
\begin{equation}
    {\cal H}_{\cal P}^{(g)\, [n]}(\hat x', \hat z)\, \delta \bigg( \frac{G(\hat x', \hat z)}{\hat x' \hat z^2} \bigg) \sim {\cal H}_{\cal P}^{(g)\, [n];\, (0)}(\hat x', \hat z)\  \delta_{\rm eff} (\hat x', \hat z)\ ,
\end{equation}
with
\begin{align}
    \delta_{\rm eff} (\hat x', \hat z) & = \hat x_{\rm max} \bigg[ \frac{1}{2}\left(\log \frac{M_\psi^2 + Q^2}{\qT^2} - 1 - \log \frac{M_\psi^2}{M_\psi^2 + Q^2}  \right)\delta(1 - \hat x')\, \delta(1 - \hat z) \nt \\
    & \phantom{=} + \frac{\hat x'}{(1 - \hat x')_+}\, \delta(1 - \hat z) + \frac{M_\psi^2 + Q^2}{M_\psi^2/\hat z + Q^2}\frac{\hat z}{(1 - \hat z)_+}\, \delta(1 - \hat x') \bigg]\ .
\label{eq: effective delta expansion}
\end{align}
Considering the contributions from the various terms in Eq.~\eqref{eq: effective delta expansion}, we found that the small-$\qT$ limit is dominated by the first two terms for this $\gamma^* g$ channel (and similarly the first two terms of Eq.~\eqref{eq: naive delta expansion} for the quark and antiquark channels).
Instead, the contribution coming from the “$+$”-distribution of $\hat z$ is subdominant and can be neglected in the following. In principle, this last term will lead to a fragmentation-like contribution to the process considered here. We will further comment on the connection to a fragmentation description below.

Hence, the leading power {behavior} of the structure functions in the ITM region is given by
\begin{equation}
\begin{aligned}
    F_{UU,{\cal P}} & = \sigma_{UU,{\cal P}}  \frac{\alpha_s}{\qT^2}\, \Big[ L(\qT^2)\, f_1^g(x;\mu^2) + \left( P_{gg} \otimes f_1^g + P_{gi} \otimes f_1^i\right)(x;\mu^2) \Big]\ , \\
    F_{UU}^{\cos 2\phi_\psi} & = \sigma_{UU}^{\cos 2\phi_\psi} \frac{\alpha_s}{\qT^2} \left( \delta P_{gg} \otimes f_1^g + \delta P_{gi} \otimes f_1^i\right)(x;\mu^2) \ ,
\label{eq: collinear structure functions at ITM}
\end{aligned}
\end{equation}
{while $F_{UU}^{\cos \phi_\psi}$ is suppressed by a factor of $|\qT|/\muH$ w.r.t. the other structure functions. This is in accordance with the TMD formula in Eq.~\eqref{eq: SIDIS diff cross section TMD} which does not show any ${\cos \phi_\psi}$ contribution too.}
The logarithmic function $L( \qT^2)$ reads
\begin{equation}
    L(\qT^2) = C_A \left(\log\frac{M_\psi^2 + Q^2}{\qT^2} - 1 - \log \frac{M_\psi^2}{M_\psi^2 + Q^2} - \frac{11 - 4\, n_f T_R/C_A}{6}\right) \ ,
\label{eq: logarithmic function}
\end{equation}
and the quantities $\sigma_{UU,{\cal P}}$ and $\sigma_{UU}^{\cos 2\phi_\psi}$ are related to the partonic process $\gamma^* g \to c\bar c[n]$ and they correspond to (see Ref.~\cite{Boer:2020:epJpsiMatching})
\begin{equation}
\begin{aligned}
    \sigma_{UU, \perp} & = \frac{e_c^2\, \alpha \alpha_s}{M_\psi \left( M_\psi^2 + Q^2 \right)} 
        \left[ \langle {\cal O}[{}^1 S_0^{[8]}] \rangle + 4 \frac{7 M_\psi^4 + 2 M_\psi^2 Q^2 +3 Q^4}{M_\psi^2 \left( M_\psi^2 + Q^2 \right)^2} \langle {\cal O}[{}^3 P_0^{[8]}] \rangle \right] \delta(1 - z) \ , \\
    \sigma_{UU, \varpar} & = \frac{e_c^2\, \alpha \alpha_s}{M_\psi \left( M_\psi^2 + Q^2 \right)}
        \left[ 16 \frac{Q^2}{ \left(M_\psi^2 + Q^2 \right)^2} \langle {\cal O}[{}^3 P_0^{[8]}] \rangle \right] \delta(1 - z) \ , \\
    \sigma_{UU}^{\cos 2\phi_\psi} & = \frac{e_c^2\, \alpha \alpha_s}{M_\psi \left( M_\psi^2 + Q^2 \right)}
        \left[ - \langle {\cal O}[{}^1 S_0^{[8]}] \rangle + 4 \frac{{3 M_\psi^2 - Q^2}}{M_\psi^2 \left( M_\psi^2 + Q^2 \right)} \langle
 {\cal O}[{}^3 P_0^{[8]}] \rangle \right] \delta(1 - z) \ .
\end{aligned}
\end{equation}
Moreover, $P_{ab}$ in Eq.~\eqref{eq: collinear structure functions at ITM} denotes the leading order, fully unpolarized splitting functions, that can be found in Ref.~\cite{Collins:2011:QCDbook}, while $\delta P_{ab}$ are the splitting function of an unpolarized parton into a linearly polarized gluon, which can be found in Refs.~\cite{Sun:2011:GluonHiggs,Catani:2010:QCDResumGluon}.
{The convolution (denoted by the “$\otimes$” symbol) between these splitting functions and the parton distribution functions is defined as
\begin{equation}
    ({\cal P} \otimes f_1^a)(x; \mu^2) = \int_x^1 \frac{\d \hat x'}{\hat x'}\, {\cal P}_{ab}(\hat x'; \mu^2)\, f_1^b(x/\hat x'; \mu^2) \ ,
\end{equation}
where ${\cal P}_{ab}$ denotes either $P_{ab}$ or $\delta P_{ab}$.}

{The logarithmic function defined in Eq.~\eqref{eq: logarithmic function} is our most important difference compared to Ref.~\cite{Boer:2020:epJpsiMatching}, where the logarithmic function contains twice the logarithm $\log [(M_\psi^2 + Q^2)/{\qT^2}]$ compared to Eq.~\eqref{eq: logarithmic function}.
This is due to the presence of the poles, not considered in Ref.~\cite{Boer:2020:epJpsiMatching}. Indeed, it is through the inclusion of Eqs.~\eqref{eq: sigma B approximation} and~\eqref{eq: sigma C approximation} that in Eq.~\eqref{eq: logarithmic function} one of the logarithms has been removed. The price to pay corresponds to the novel $\qT$-independent terms found, namely $1 + \log[(M_\psi^2 + Q^2)/{M_\psi^2}]$. Clearly, Eq.~\eqref{eq: logarithmic function} has an impact on the TMDShF derivation too, as will be discussed in Sec.~\ref{sec: matching Low -> Int}.}
{Besides,} Eq.~\eqref{eq: logarithmic function} implies the presence of divergences related to soft gluon emission from the leading order $\gamma^* g \to c\bar c[n]$ process. It is {then} possible to check the validity of this expression by investigating the soft-limit of Eq.~\eqref{eq: partonic interaction HTM} via the eikonal method, as done in Appendix~\ref{app: eikonal method}.

{Although our work is based on the $J/\psi$ production in SIDIS, we expect that the presence of the poles as in Eq.~\eqref{eq: structure functions Laurent expansion} is an intrinsic feature of any inclusive quarkonium production, and they apply to different processes and observables too. Hence, the suppression of the $\qT$-logarithm in Eq.~\eqref{eq: logarithmic function} is not an exclusive outcome of the specific process under consideration, but rather a general statement.
Thus, these discontinuities may be connected to other regularization procedures associated with CO contributions to heavy quarkonium productions. While it is worthwhile to further pursue these connections in the NRQCD factorization, we consider such a study to be beyond the scope of the current paper but we hope to address it in the future.
However, to emphasize the importance of further investigation, we will briefly comment on the similarities of our findings with those obtained by adopting the fragmentation function description.
}

The same cross section in the HTM region can be expressed in terms of fragmentation functions, as shown in Refs.~\cite{Kang:2014pya,Ma:2014:OniapTexpansion,Kang:2014tta}. 
Hence, the TMDShF may also be seen as a fragmentation-like function of a $c \bar c$ into a $J/\psi$ evaluated at ITM. The evolution of the latter has been studied in Ref.~\cite{Ma:2014:OniapTexpansion},  which includes real contributions having a component proportional to the $(1 - \hat z)_+$  distribution and another one to $\delta(1 - \hat z)$. Hence, our analysis is related to the latter term.
However, an important difference concerns the integration range of the outgoing gluon. The integration of the soft-gluon momentum in our case has a lower limit set by the $J/\psi$ transverse momentum (see Eq.~\eqref{eq: x_g range} of Appendix~\ref{app: eikonal method}), whereas no lower limit is present in Ref.~\cite{Ma:2014:OniapTexpansion} causing infrared divergences.
Therefore, the connection between our work and the fragmentation-function description cannot be carried out further without the inclusion of next order (real and virtual) contributions and{, as previously stated, } we leave this discussion to further studies.

\subsection{From low to intermediate transverse momentum}
\label{sec: matching Low -> Int}
In this section we evaluate the evolution of the structure functions valid in the LTM region ({$|\qT| \ll \muH$}) up to the ITM region. Even if not formally proven, there are strong arguments in {favor} of TMD factorization \cite{Bacchetta:2018:GluonTMDsEIC}. Therefore, the differential cross section for the semi-inclusive production of a $J/\psi$ with a small transverse momentum component is given by Eq.~\eqref{eq: SIDIS diff cross section TMD}. In this case, the structure functions ${\cal F}$ can be calculated from the partonic process
\begin{equation}
    \gamma^* \, (q) + g \, (p_a) \to c\bar c[n]\, (P_\psi)\ ,
\label{eq: partonic interaction LTM}
\end{equation}
where, contrarily to the HTM case, the initial gluon has a non-negligible transverse momentum component w.r.t.\ the parent proton, namely
\begin{equation}
    p_a^\mu = \xi P^\mu + \pTfv^\mu\ ,
\end{equation}
with $\pTfv^2 = - \pT^2$.
Hence, Eq.~\eqref{eq: partonic interaction LTM} leads to 
\begin{equation}
\begin{aligned}
    {\cal F}_{UU, \perp} & = 2 \pi^2 \frac{e_c^2\, \alpha \alpha_s}{M_\psi \left( M_\psi^2 + Q^2 \right)} \left( {\cal C} \left[ f_1^g\, \Delta^{[{}^1 S_0^{[8]}]} \right] + 4 \frac{7 M_\psi^4 + 2 M_\psi^2 Q^2 +3 Q^4}{M_\psi^2 \left( M_\psi^2 + Q^2 \right)^2}\, {\cal C} \left[ f_1^g\, \Delta^{[{}^3 P_0^{[8]}]} \right] \right) \ , \\
    {\cal F}_{UU, \varpar} & = 2 \pi^2 \frac{e_c^2\, \alpha \alpha_s}{M_\psi \left( M_\psi^2 + Q^2 \right)} \left( 16 \frac{Q^2}{ \left(M_\psi^2 + Q^2 \right)^2}\, {\cal C} \left[ f_1^g\, \Delta^{[{}^3 P_0^{[8]}]} \right] \right) \ , \\
    {\cal F}_{UU}^{\cos  2 \phi_\psi } & = \frac{\pi^2}{2} \frac{e_c^2\, \alpha \alpha_s}{M_\psi \left( M_\psi^2 + Q^2 \right)} \left( - \, {\cal C} \left[ w\, h_1^{\perp g}\, \Delta_h^{[{}^1 S_0^{[8]}]} \right] + 4 \frac{{3 M_\psi^2 - Q^2}}{M_\psi^2 \left( M_\psi^2 + Q^2 \right)} \, {\cal C} \left[ w\, h_1^{\perp g}\, \Delta_h^{[{}^3 P_0^{[8]}]} \right] \right) \ .
\label{eq: TMD structure functions}
\end{aligned}
\end{equation}
Following Refs.~\cite{Echevarria:2019:TMDShF, Fleming:2019:TMDShF, Boer:2020:epJpsiMatching,DAlesio:2021:epJpsiMatchingPol, Bor:2022:EvolTMDShF}, we consider from the beginning the TMD factorized formula that includes the presence of a TMDShF, $\Delta^{[n]}$ or $\Delta_h^{[n]}$, which is related to the production of a $J/\psi$ with a small transverse momentum component w.r.t.\ the photon and proton.
As commented below Eq.~\eqref{eq: effective delta expansion}, in this paper we focus on the $\delta(1 - \hat{z})$ contribution from the TMDShF. Subdominant terms and higher order corrections are also expected to contribute away from $z = 1$. In that case the description of the heavy quark pair that hadronizes into the heavy quarkonium state will be even more similar to a single-parton TMD fragmentation functions description applied in light hadron production. See also Ref.~\cite{Lee:2021oqr} for a description involving both single parton and parton-pair fragmentation processes.  

In the above equations convolutions between the TMD distributions and the TMDShFs appear, namely
\begin{equation}
\begin{aligned}
    {\cal C}\left[ f_1^g\, \Delta^{[n]} \right](x, z, \qT^2) & = \int \d^{2}\pT \int \d^{2}\kT \,  
    \delta^{2}(\pT+\kT-\qT) \, 
        f_{1}^{g}(x,\pT^{2})\, \Delta^{[n]}(z,\kT^{2}) \ , \\
    {\cal C}\left[w\, h_1^{\perp g}\, \Delta^{[n]}_h \right] (x, z, \qT^2) & = \int \d^{2}\pT \int \d^{2}\kT \,  
    \delta^{2}(\pT+\kT-\qT) \, 
        w(\pT, \kT)\, h_1^{\perp g}(x,\pT^{2})\, \Delta^{[n]}_h(z,\kT^{2})\ ,
\label{eq: TMD convolutions}
\end{aligned}
\end{equation}
where in the last line we introduced the weight function $w$ defined as (with $M_p$ the proton mass)
\begin{equation}
    w (\pT, \kT) = \frac{1}{M_p^2 \left(\pT + \kT \right)^2} \Big[ 2 (\pT \cdot \kT)^2 + \pT^2 \left(\pT^2 - \kT^2 \right) \Big]\ .
\end{equation}

Beyond the parton model approximation, soft gluon radiation to all orders is included into an exponential Sudakov factor. One can relate its logarithmic divergences to the TMD objects (both PDFs and shape function) involved in the reactions, whereas the remaining perturbative $\qTfv$-independent corrections are collected into the hard term. 
As a consequence of the regularization of their ultraviolet and rapidity divergences, TMD-PDFs depend on two different scales, respectively $\mu$ and $\sqrt \zeta$. We take these two scales to be equal and denote them by $\mu$.
In contrast, there are no rapidity divergences associated to the TMDShF. Thus, we can impose for its rapidity parameter $\zeta_{\Delta} = 1$, in line with Ref.~\cite{delCastillo:2021:TMDdijet/diOniaEIC}.

Implementing TMD evolution is more easily done in impact parameter space, where convolutions in the cross section become simple products. Besides, in Ref.~\cite{Boer:2020:epJpsiMatching} it was found that, up to the precision considered, from the matching procedure it is possible to deduce only the naive order-$\alpha_s^0$ part of $\Delta_h^{[n]}(z, \kT^{2})$ that is proportional to $\delta(\kT^2)$. Note that in reality smearing effects will be involved, but that small-$\kT$ behavior cannot be obtained from a perturbative matching calculation, at least up to the perturbative order considered here. As a consequence, in the following we focus on the convolution ${\cal C}[f_{1}^{g}\Delta^{[n]}]$. 

We define the Fourier transform of {$f_{1}^{a}(x,\bm{p}_{T}^{2})$} as
\begin{equation}
\widetilde f_{1}^{a}(x,\bT^{2}) = \frac{1}{2\pi} \int \d^{2}\pT \, \e^{i \bT \cdot \pT} \,f_{1}^{a}(x,\pT^{2}) \ ,
\label{eq: PDF FT}
\end{equation}
and the Fourier transformed TMDShF as
\begin{equation}
\widetilde \Delta^{[n]}(z, \bT^{2}) = \frac{1}{2\pi} \int \d^{2}\kT \, \e^{i \bT \cdot \kT} \, \Delta^{[n]}(z, \kT^{2}) \ ,
\label{eq: PDF SF}
\end{equation}
from which 
\begin{equation}
{\cal C}[f_{1}^{g}\,\Delta^{[n]}](x, z, \qT^{2}; \mu^2 = \muH^2) = \int \d^{2}\bT \, \e^{-i\bT \cdot \qT} \widetilde f_1^g(x, \bT^2; \muH^2) \,  \widetilde{\Delta}^{[n]}(z, \bT^{2}; \muH^2)\ ,
\label{eq: TMD convolution in bT space}
\end{equation}
where we fixed the factorization scale so that the convolutions are evaluated at the hard scale.
The perturbative tail of the fully unpolarized gluon TMD $f_1^g$, valid in the limit $|\bT| \ll 1/\Lambda_{\rm QCD}$, is given by \cite{Collins:2011:QCDbook}
\begin{equation}
\widetilde f_{1}^{g}(x,\bT^{2}; \muH^2) = \frac{1}{2\pi} \sum_{a = q,\bar q,g}(C_{g/a}\otimes {f_1^{a}})(x;\mu_{b}^2)\, \e^{-\frac{1}{2} S_{A}^{g}(\bT^2; \muH^2)} \ ,
\label{eq: TMDPDF tail}
\end{equation}
where $\mu_{b} = b_{0}/|\bT|$ with $b_{0} = 2 \e^{-\gamma_{\scriptscriptstyle E}}\approx 1.123$.
Note that the coefficient function $C_{a/b}$ in Eq.~\eqref{eq: TMDPDF tail} can be expanded in powers of $\alpha_s$
\begin{equation}
    C_{g/a}(x,\mu_{b}^2) = \delta_{ab}\, \delta(1 - x) + \sum_{k = 1}^{\infty}C_{g/a}^{(k)}(x)\bigg(\frac{\alpha_{s}(\mu_{b})}{\pi}\bigg)^{k}\ ,
\label{eq: coefficient functions}
\end{equation}
and can be explicitly found in Ref.~\cite{Echevarria:2015:EvolTMDHiggs, Collins:2011:QCDbook}.
Nevertheless, the coefficient $C_{g/a}^{(k)}$ in the right-hand side of Eq.~\eqref{eq: coefficient functions} will not enter in the following (leading order) discussion since they are independent of the parameter $\bTfv$. Consequently, their explicit expression at all orders is not required. Furthermore, the (leading order) Sudakov factor $S_A$ present in Eq.~\eqref{eq: TMDPDF tail} reads
\begin{align}
    S_A^{g}(\bT^2; \muH^2) 
    & = \frac{C_A}{\pi} 
        \int_{\mu_{b}^{2}}^{ \muH^2} \frac{\d\mu'^{2}}{\mu'^{2}} \, \alpha_s(\mu') \, \bigg[\log \frac{\mu^{2}}{{\mu'}^{2}} - \frac{11 - 4 n_{f} T_{R}/C_A}{6} \bigg] \nt \\
    & = \frac{C_A}{\pi}
        \alpha_{s} \bigg(\frac{1}{2} \log^{2}\, \frac{\muH^2}{\mu_{b}^2} - \frac{11 - 4 n_{f} T_{R}/C_A}{6} \log \frac{\muH^2}{\mu_{b}^2}\bigg)\ ,
\label{eq: perturbative Sudakov}
\end{align}
where in the last line the running of the coupling has been neglected. By inserting Eqs.~\eqref{eq: coefficient functions} and~\eqref{eq: perturbative Sudakov} in Eq.~\eqref{eq: TMDPDF tail} and using the DGLAP equations to evolve the PDF from a scale $\muH$ down to the scale $\mu_b < \muH$, we find that up to order $\alpha_s$ the perturbative tail of the gluon TMD-PDF reads~\cite{Echevarria:2015:EvolTMDHiggs}
\begin{align}
\widetilde{f}_{1}^{g}(x,\bT^{2}; \muH^2) & = 
    \frac{1}{2\pi} \bigg\{
        {f_1^{g}}(x;\muH^2)-\frac{\alpha_s}{2\pi}\bigg[
        C_A\bigg(\frac{1}{2}\,\log^{2}\, \frac{\muH^2}{\mu_{b}^2} - \frac{11- 4\, n_{f} T_{R}/C_A}{6}\log\frac{\muH^2}{\mu_{b}^2}\bigg)
        {f_1^{g}}(x; \muH^2) \nt \\
    & \phantom{=} + \left(P_{gg}\otimes {f_1^{g}} + P_{gi}\otimes {f_1^{i}} \right)(x;\muH^2)\, \log \frac{\muH^2}{\mu_{b}^2} - 2 \sum_{a = q,\bar{q},g}(C_{g/a}^{(1)}\otimes {f_1^{a}})(x;\muH^2)\bigg] \bigg\}\ ,
\label{eq: final FT TMD}
\end{align}
where once again $P_{ab}$ denotes the leading order splitting functions~\cite{Collins:2011:QCDbook}.
Employing this, and by requiring that the TMD expressions evolved to the scale $\muH^2 \equiv \widetilde Q^2 = M_\psi^2 + Q^2$ match with the expansion of the collinear ones obtained in Eq.~\eqref{eq: collinear structure functions at ITM}, we deduce the TMDShF perturbative tail 
\begin{align}
\widetilde{\Delta}^{[n]}(z, \bT^2; \mu^2 = \widetilde Q^2) & = 
    \frac{1}{2\pi} \left[ 1 + \frac{\alpha_{s}}{2\pi} \,C_{A} \left( 1 + \log \frac{M_\psi^2}{M_\psi^2 + Q^2} \right) \log \frac{\widetilde Q^{2}}{\mu_{b}^{2}} \right] \langle {\cal O}[n] \rangle \, \delta(1 - z) \nt \\ & \phantom{=}
        + {\cal O}(\alpha_{s}^{2}) + {\cal O} (\bTfv \Lambda_{\rm QCD})\ ,
\label{eq: SF in bT space}
\end{align}
which in momentum space becomes
\begin{equation}
{\Delta}^{[n]}(z, \kT^{2}; \widetilde Q^2) = - \frac{\alpha_{s}}{2\pi^{2}\kT^{2}} \,C_{A}  \left( 1 + \log \frac{M_\psi^2}{M_\psi^2 + Q^2} \right) \langle {\cal O}[n] \rangle \, \delta(1 - z) \ ,
\label{eq: SF perturbative tail}
\end{equation} 
valid in the $|\kT| \gg \Lambda_{\rm QCD}$ limit.
Inserting Eq.~\eqref{eq: SF in bT space} in Eq.~\eqref{eq: TMD convolution in bT space}, we find that the convolution in momentum space is given by
\begin{align}
{\cal C} [f_{1}^{g}\,\Delta^{[n]}](x, z, \qT^{2}; \widetilde Q^2)
    & = \frac{\alpha_s}{2\pi^{2}\qT^{2}} \Big[L(\qT^2)\, {f_1^{g}}(x;\widetilde Q^2) + \big( P_{gg}\otimes {f_1^{g}} + P_{gi}\otimes {f_1^{i}}  \big)(x;\widetilde Q^2)\Big] \langle {\cal O}[n] \rangle \, \delta(1 - z)\ ,
\end{align}
where $L(\qT^2)$ is the logarithmic function defined in Eq.~\eqref{eq: logarithmic function}.
Hence, with the choice $\mu = \widetilde Q$, the first two lines of Eq.~\eqref{eq: TMD structure functions} and the first line of Eq.~\eqref{eq: collinear structure functions at ITM} match.
{Note how the modification of Eq.~\eqref{eq: logarithmic function} compared to Ref.~\cite{Boer:2020:epJpsiMatching} has a significant impact on the TMDShF expression. Indeed, the TMDShF perturbative tail in Eq.~\eqref{eq: SF perturbative tail} does not contain any kind of logarithmic divergence in $\kT$, being tamed by the presence of the heavy mass. We emphasized that the absence of $\kT$-divergent terms associated to the quarkonium is in accordance with other works in the literature, {e.g.}~Refs.~\cite{Sun:2012:ppJpsiResum, Zhu:2012:OpenTopHadronCollider, Zhu:2013:epOpenQuark, Echevarria:2019:TMDShF, Fleming:2019:TMDShF, Echevarria:2022:talk@Tansversity}.}

For completeness, we remark that the matching of $F_{UU}^{\cos 2\phi_\psi}$ and ${\cal F}_{UU}^{\cos 2\phi_\psi}$, which involves the second convolution in Eq.~\eqref{eq: TMD convolutions}, is fulfilled by taking the perturbative tail of $h_1^{\perp g}$ \cite{Sun:2011:GluonHiggs} up to $\alpha_s$ order
\begin{equation}
    h_1^{\perp g}(x, \pT^2) = \frac{\alpha_s}{\pi^2} \frac{2\, {M_p^2}}{\pT^4} \big( \delta P_{gg} \otimes f_1^g + \delta P_{gi} \otimes f_1^i \big)(x)  + {\cal O}(\alpha_s^2) 
\end{equation}
and the leading order naive shape function
\begin{equation}
    \Delta_h^{[n]} (z, \kT^2) = \delta(\kT^2)\, \langle {\cal O}[n] \rangle\, \delta(1 - z)  + {\cal O}(\alpha_s) \ ,
\end{equation}
from which
\begin{equation}
    {\cal C} [w\, h_{1}^{\perp g}\,\Delta_h^{[n]}](x, z, \qT^{2}; \mu^2) = \frac{2}{\pi^2} \frac{\alpha_s}{\qT^2} \big( \delta P_{gg} \otimes f_1^g + \delta P_{gi} \otimes f_1^i \big)(x; \mu^2)\, \langle {\cal O}[n] \rangle \, \delta(1 - z) \ .
\end{equation}
Since the $h_1^{\perp g}$ expansion starts at order $\alpha_s$, we notice that to get the non-trivial perturbative tail of $\Delta_h$ it is required that the SIDIS cross section within NRQCD is evaluated at order $\alpha \alpha_s^3$. However, this calculation is currently unavailable.

\section{Universality}
\label{sec: Universality}
In the previous section, we found that an extra factor $\Delta$ is needed to absorb all the $\qT$-divergent terms coming from the collinear limit, and we identified it as the dominant TMDShF perturbative tail. However, it has been obtained at the particular scale $\widetilde Q$, whereas for more general application it needs to be considered at a general scale $\muH$.
This can be obtained by tracing back the $\muH$ dependence in Eq.~\eqref{eq: logarithmic function}, that 
is related to the full Sudakov factor for $J/\psi$ production in SIDIS in terms of this general scale and up to order $\alpha_s$, namely
\begin{equation}
    S^{ep,\psi}_{A}(\bT^2; \muH^2) = \frac{1}{2} S_A^g(\bT^2; \muH^2) + B_{ep}(\muH^2)\log \frac{\muH^2}{\mu_{b}^2}\ ,
\label{eq: Onia Sudakov ep}
\end{equation}
where
\begin{equation}
    B_{ep}(\muH^2) = - \frac{\alpha_s}{2 \pi} C_A \left(1 + \log\frac{M_\psi^2\, \muH^2}{\big( M_\psi^2 + Q^2 \big)^2} \right) \ .
\label{eq: B term ep}
\end{equation}    
We checked that Eq.~\eqref{eq: Onia Sudakov ep} (and subsequently Eq.~\eqref{eq: B term ep}) agrees in the kinematic limit corresponding to a bound pair with the Sudakov factor obtained in the open heavy-quark pair production in electron-proton collisions, which can be found in Ref.~\cite{Zhu:2013:epOpenQuark}.

It is not natural to fully include Eq.~\eqref{eq: B term ep} into something that we identify as the TMDShF. 
Indeed, being a quarkonium-related object, its complete dependence is given by $\Delta_{\rm ShF}^{[n]}(z, \kT^2; M_\psi^2, \mu^2 = \muH^2)$, while it may depend on the process-related hard-quantity $Q$ only via the $\muH$ choice.
Thus, the $Q^2$ dependence deriving from Eq.~\eqref{eq: B term ep} must stem from a process dependent part, which can be incorporated into an extra process-dependent factor $S(\bT^2; M_\psi^2, Q^2, {\mu^2 = \muH^2})$.

Therefore, we split the full $\Delta^{[n]}_{ep}$ into these two terms:\footnote{
Here we introduced the subscript “$ep$” to underline that this $\Delta$ has been obtained for SIDIS.
}

\begin{equation}
    \Delta^{[n]}_{ep}(\muH^2) = \Delta_{\rm ShF}^{[n]}(\muH^2) \times S_{ep}(\muH^2)\ .
\end{equation}
The $\Delta_{\rm ShF}^{[n]}$ is what we truly identify as the TMDShF and is universal because it solely depends on $M_\psi$. 
Instead, the $S_{ep}$ is an extra soft factor which incorporates the specific process dependence and it can be removed by a proper choice of the factorization scale $\mu = \muH$. 
This implies that at that scale the full $\Delta^{[n]}_{ep}$ is equivalent to the TMDShF.
At this level, the simplest way to perform the splitting in $\bTfv$-space is to take
\begin{align}
    \widetilde \Delta^{[n]}_{\rm ShF} (z, \bT^2; \muH^2) & = \frac{1}{2\pi} \left[ 1 + \frac{\alpha_s}{2\pi} C_A \left( 1 + \log \frac{M_\psi^2}{\muH^2}\right) \log \frac{\muH^2}{\mu_b^2} \right] \langle {\cal O}[n] \rangle\, \delta(1 - z)\ , 
    \label{eq: universal TMDShF}\\
    S_{ep} (\bT^2; \muH^2) & = 1 + \frac{\alpha_s}{2\pi} C_A  \left(2 \log \frac{\muH^2}{M_\psi^2 + Q^2} \right)\log\frac{\muH^2}{\mu_b^2} \ .
\end{align}
With this splitting convention and by taking $\muH \equiv \widetilde Q$, the full $\Delta^{[n]}_{ep}$ reduces to the TMDShF, implying that the latter is given by Eq.~\eqref{eq: SF perturbative tail}.

To test the proposed factorization, one may consider another process and check if it is possible to identify the same TMDShF in Eq.~\eqref{eq: universal TMDShF}.
We take into account $J/\psi$ production in hadron collisions, namely $pp \to J/\psi + X$.\footnote{
It should be mentioned that a $J/\psi$ produced from $gg$ fusion is necessarily in the CO state, because production of a massive CS vector state from two massless gluons is not possible (Landau-Yang theorem). Nevertheless, in case of a CO final state in $pp$ scattering the gluon TMD will involve a different gauge link structure than in $ep$ and TMD factorization may not even hold. As there is much unclear about this, we will ignore this complicating matter in this work.}
For this process the small-$\qT$ {behavior} of the cross section evaluated in the HTM region has been calculated in Ref.~\cite{Sun:2012:ppJpsiResum}.
The corresponding Sudakov factor can be written as
\begin{equation}
    S^{pp,\psi}_{A}(\bT^2; \muH^2) = 
    S_A^g(\bT^2; \muH^2) + B_{pp}(\muH^2) \log \frac{\muH^2}{\mu_{b}^2} \ ,
\label{eq: Onia Sudakov}
\end{equation}
where
\begin{equation}
    B_{pp}(\muH^2) = - \frac{\alpha_s}{2 \pi} C_A \left( 1 + 2 \log\frac{\muH^2}{M_\psi^2} \right) \ ,
\label{eq: B term pp}
\end{equation}
in which the first term of Eq.~\eqref{eq: B term pp} is directly related to the $\delta_{8c}$ term in Ref.~\cite{Sun:2012:ppJpsiResum}.
Also in this case we checked that previous equations agree in the kinematic limit corresponding to a bound pair with the open heavy-quark pair production Sudakov factor, which can be found in the literature (for instance Ref.~\cite{Zhu:2012:OpenTopHadronCollider}).
Moreover, even if it is possible to produce quarkonia in a CS state ({e.g.}~$\eta_c$), for $pp$ our perturbative tail only applies to CO states. Despite this, we cannot exclude that a non trivial TMDShF perturbative tail applies to the CS channel too, if one goes to next orders in perturbation theory.

Although the full $\Delta_{pp}^{[n]}$ is different from $\Delta_{ep}^{[n]}$, we can still identify the same $\Delta_{\rm ShF}^{[n]}$ in Eq.~\eqref{eq: universal TMDShF}, which is now combined with a different (extra) soft factor $S_{pp}$, namely
\begin{equation}
    \Delta^{[n]}_{pp}(\muH^2) = \Delta_{\rm ShF}^{[n]}(\muH^2) \times S_{pp}(\muH^2)\ ,
\end{equation}
with
\begin{equation}
    S_{pp}(\muH^2) = 1 + \frac{\alpha_s}{2 \pi} C_A \left( 3 \log \frac{\muH^2}{ M_\psi^2} \right) \log\frac{\mu^2}{\mu_b^2}\ .
\end{equation}
Interestingly, for $S_{pp}$ the coefficient in front of the $\log$ is “$3$”, whereas the same coefficient for $S_{ep}$ is “$2$”, which corresponds to the number of TMD quantities (PDFs and shape functions) involved. Hence, even if process dependent, these terms are the same apart from the number of TMDs involved. This may allow to guess the required term for other processes, such as for di-quarkonium production in $pp$ collisions (if that factorizes at all for CO-CO production). 

The factor $S_{pp}$ reduces to $1$ when $\muH = M_\psi$, such that $\Delta^{[n]}_{pp}(M_\psi^2) = \Delta^{[n]}_{\rm ShF}(M_\psi^2)$. 
For this scale choice, $\Delta^{[n]}_{\rm ShF}(M_\psi^2)$ is compatible with the corresponding one presented in Ref.~\cite{Fleming:2019:TMDShF} for $\chi_c$ decay into light-quarks, where the NLO TMDShF up to corrections of ${\cal O}(|\kT|^{-1})$ is given by a constant too.

According to our findings, in principle one may obtain the value of $\Delta^{[n]}_{ep}(\widetilde Q^2)$ from the experimentally determined $\Delta^{[n]}_{pp}(M_\psi^2) = \Delta^{[n]}_{\rm ShF}(M_\psi^2) \, \big(\!\neq \Delta^{[n]}_{ep}(M_\psi^2)\big)$, by evolving $\Delta^{[n]}_{\rm ShF}(M_\psi^2)$ to $\Delta^{[n]}_{\rm ShF}(\widetilde Q^2)$.
Hence, we propose a strategy for the extraction of the TMDShF from different processes, relying on their factorizability. For processes where we have a dominant hard scale it is reasonable to expect that by setting $\muH$ equivalent to it we reduce our uncertainties in the extraction of the TMDShF.\footnote{
This applies to both $pp$, where we have only $M_\psi$, and SIDIS, if $Q \gg M_\psi$ or $Q \ll M_\psi$ (including photoproduction).}
Then, this term can be re-used for every process involving $J/\psi$ by evolving $\Delta^{[n]}_{\rm ShF}$ to the scale $\muH'$ and combining it with the proper process-dependent extra soft factor $S(\muH^{\prime\, 2})$. 

{For completeness, we mention that the soft factor derived for the open heavy-quark pair production also involves an additional process-dependent factor~\cite{Catani:2014:ppOpenQuark, Catani:2021cbl, Ju:2022wia} (which is sometimes denoted by $\Delta$, but should not to be confused with ours).}
This additional factor stems from soft radiation in the $Q \bar Q$ production and can in principle even depend on the angle of $\qT$. Hence, it is natural to expect an additional process-dependent soft term in the quarkonium case too. In that sense we expect that our extra soft term $S$ will acquire azimuthal and rapidity dependences if one goes beyond the order and approximation we have considered, as they are present in the $\Delta$ quantity of {Refs.~\cite{Catani:2014:ppOpenQuark, Ju:2022wia}}.

\section{Conclusions}
\label{sec: Conclusions}
In this work, we revised the procedure to derive the leading order TMDShF perturbative tail for heavy quarkonium production.
We focused on the SIDIS unpolarized cross section, which is parameterized in terms of structure functions.
In particular, we considered the cross section evaluated at low $\qTfv$ and order $\alpha\alpha_s$, which involves the convolution between the gluon TMD-PDF and a general TMDShF, taking the reasonable assumption that factorization holds.
This description should match the collinear one at high $\qTfv$ and order $\alpha\alpha_s^2$ when both are evaluated at intermediate $\qTfv$, namely $\Lambda_{\rm QCD} \ll |\qT| \ll \muH$.
{We emphasize that, although the exact choice of $\muH$ is important from a phenomenological point of view where it is advantageous to extend the intermediate-$\qT$ region, our findings hold for any choice of $\muH$.}

We show that in the high transverse momentum region, these structure functions present poles when the small-$\qT$ limit is taken. 
{We expect that these poles will be contained in other hard amplitudes concerning inclusive quarkonia production.}
{Therefore, we presented} a systematic way to deal with them, showing how they provide non-negligible terms in the expansion {at small $\qT$}. These terms, neglected in \cite{Boer:2020:epJpsiMatching,DAlesio:2021:epJpsiMatchingPol}, significantly alter our findings of the TMDShF perturbative tail. 
At variance with previous works, it does not present a logarithmic dependence on the transverse momentum (double-logarithm in $\bT$-space), which makes them different from usual TMD fragmentation functions for light hadron production.
However, this non-logarithmic dependence is in agreement with other works~\cite{Echevarria:2019:TMDShF, Fleming:2019:TMDShF}, and with the Sudakov factors obtained for open heavy-quark pair production in electron-proton and proton-proton collisions. 

We remark that our results on the transverse momentum dependence of the TMDShFs hold for every CO quarkonium state with the same quantum numbers as the $J/\psi$ we considered, {e.g.}~$\Upsilon(nS)$ and $\psi(2S)$. The magnitude of TMDShFs can be different though and is determined by the LDMEs.
This conclusion holds up to the precision considered, corresponding to the $\alpha\alpha_s^2$ and $v^4$ orders in the NRQCD double expansion.
Moreover, the same considerations apply if we take into account the polarization of the $J/\psi$, since the kinematics is the same.
Namely, we have the same TMDShF perturbative tail for both the longitudinal and transverse $J/\psi$ polarization states.
Besides, to check that the same form of the TMDShF applies for observables involving $h_1^{\perp g}$ we would require the computation of the cross section within NRQCD at higher order in $\alpha_s$, both for polarized and unpolarized $J/\psi$ productions.
However, this calculation is still unavailable.

Furthermore, we showed that if we consider the evolution w.r.t.\ the factorization scale $\mu$, the TMDShFs would have to depend on the hard scale $Q$ too. As it is not reasonable to include this dependence into a quantity that is related to the quarkonium formation solely, we considered a split into two terms: a process-independent quantity that we identify as the universal TMDShF, and an extra process-dependent soft factor. This then allows to make a connection between $ep$ and $pp$ processes, without losing predictability completely. It is also in line with results for open heavy quark production, where extra process dependent soft factors are also required, at least in $pp$ collisions {\cite{Catani:2014:ppOpenQuark, Ju:2022wia}}. 

Despite the process dependence, we showed that it is possible to extract the universal TMDShFs by appropriate choices of scales, which allows to relate different processes.
Hence, we expect that with the upcoming Electron-Ion Collider and more data provided by $pp$ facilities ({e.g.}~LHC in fixed target mode) extractions of the TMDShFs will become available in the future and new features of heavy quarkonium production will be uncovered.

\section*{Acknowledgements}
We thank Miguel Echevarría for helpful discussions and feedback.
This project has received funding from the European Union’s Horizon 2020 research and innovation programme under grant agreement No.~$824093$ (STRONG 2020) and is part of its JRA4-TMD-neXt Work-Package. 
This project has also received funding from the French Agence Nationale de la Recherche via the grant ANR-20-CE31-0015 (“PrecisOnium”) and was also partly supported by the French CNRS via the IN2P3 project GLUE@NLO.
C.P. also acknowledges financial support by Fondazione di Sardegna under the project “Proton tomography at the LHC”, project number~ F72F20000220007 (University of Cagliari). This project has also received funding from the LDRD program of LBNL, and the U.S. Department of Energy, Office of Science, Office of Nuclear Physics, under contract numbers DE-AC02-05CH11231.

\appendix
\section{The additional terms of the effective delta}
\label{app: poles}
In this appendix we provide more details on the derivation of Eq.~\eqref{eq: effective delta expansion}.
Via the Laurent expansion in Eq.~\eqref{eq: structure functions Laurent expansion} we obtained the three terms presented in Eq.~\eqref{eq: differential cross section separation}.
The first integral, $\d \sigma_A$, involves only finite terms in the double limit $\hat x', \hat z \to 1$. On the contrary, $\d \sigma_B$ and $\d \sigma_C$ include indeterminate terms. 
These are given by the poles, while other quantities can be Taylor expanded around $\hat z = 1$ and $\hat x' = 1$; {e.g.}~the quantities ${\cal H}_{\cal P}^{[n]; (k)}(\hat z)$ are decomposed as
\begin{equation}
    {\cal H}_{\cal P}^{[n]; (k)}(\hat z) = {\cal H}_{\cal P}^{[n]; (k)}(1) + \sum_m {(1 - \hat z)^m} \left. \frac{\d^m {\cal H}_{\cal P}^{[n]; (k)}(\hat z)}{\d \hat z^m} \right|_{\hat z = 1}\, .
\end{equation}
After the first order, the presence of power of {$(1 - \hat z)^m$} solves the indeterminacy, making the quantity $\frac{(1 - \hat z)^{1 + m}}{1 - \hat x}$ null in the double limit.
{Hence}, one can approximate $\hat z = 1$ and $\hat x' = 1$ whenever possible, and subsequently perform the analytic integral.
{To achieve so, we can utilize the solution of $\hat x'$ imposed by ${\delta\big( {G(\hat x', \hat z)} \big)}$, namely
\begin{equation}
	\hat x'_0 = 1 - \frac{(1 - \hat z)^2 M_\psi^2 + \hat z^2 \qT^2}{(1 - \hat z) (M_\psi^2 + \hat z Q^2) + \hat z\, \qT^2} \ .
\label{eq: x prime solution}
\end{equation}
Hence, via Eq.~\eqref{eq: x prime solution} we are able to rewrite the denominator of the poles and, subsequently, integrate analytically the remaining function. Explicitly, we have that}
\begin{align}
    \d \sigma_{\scriptscriptstyle B} & \approx \frac{\alpha}{4 y Q^2}  f^g (x; \mu^2) \sum_n \left[ \Big(1 + (1 - y)^2 \Big) \, {\cal H}_{\perp}^{(g)\, [n];\, (1)}(1) + 4\, (1 - y)\, {\cal H}_{\varpar}^{(g)\, [n];\, (1)}(1) \right] \langle {\cal O}[n] \rangle \nt \\ & \phantom{=} \times \int_{0}^{1} {\d \hat x'} \int_0^1 {\d \hat z}\, \left( \frac{1 - \hat z}{1 - \hat x^\prime} \right)\delta\big( {G(\hat x', \hat z)} \big) \, \delta(1 - z) \nt \\
    & = \frac{\alpha}{4 y Q^2}  f^g (x; \mu^2) \sum_n \left[ \Big(1 + (1 - y)^2 \Big) \, {\cal H}_{\perp}^{(g)\, [n];\, (1)}(1) + 4\, (1 - y)\, {\cal H}_{\varpar}^{(g)\, [n];\, (1)}(1) \right] \langle {\cal O}[n] \rangle \nt \\ & \phantom{=} \times Q^{2}
    \int_{0}^{1} {\d \hat x'} \int_0^1 {\d \hat z}\, \left( \frac{1 - \hat z}{{(1 - \hat z)^2 M_\psi^2 + \hat z^2 \qT^2}} \right) {\delta(\hat x' - \hat x'_0 )} \, \delta(1 - z)\nt \\
    & = \frac{\alpha}{4 y Q^2} 
    f^g (x; \mu^2)\sum_n \left[ \Big(1 + (1 - y)^2 \Big) \, {\cal H}_{\perp}^{(g)\, [n];\, (1)}(1) + 4\, (1 - y)\, {\cal H}_{\varpar}^{(g)\, [n];\, (1)}(1) \right] \langle {\cal O}[n] \rangle
    \nt \\ & \phantom{=} \times  
    \left( \frac{Q^2}{2 M_\psi^2} \log \frac{M_\psi^2}{\qT^2} \right)  \delta(1 - z) \nt \\
    & = \frac{\alpha}{4 y Q^2}  f^g (x; \mu^2)\sum_n \left[ \Big(1 + (1 - y)^2 \Big) \, {\cal H}_{\perp}^{(g)\, [n];\, (0)}(1,1) + 4\, (1 - y)\, {\cal H}_{\varpar}^{(g)\, [n];\, (0)}(1,1) \right] \langle {\cal O}[n] \rangle
    \nt \\ & \phantom{=} \times 
    \left( - \hat x_{\rm max}\, \log \frac{M_\psi^2}{\qT^2} \right) \delta(1 - z)\ , 
\label{eq: dsigma_B derivation}
\end{align}
and 
\begin{align}
    \d \sigma_{\scriptscriptstyle C} & \approx \frac{\alpha}{4 y Q^2} f^g (x; \mu^2)\sum_n \left[ \Big(1 + (1 - y)^2 \Big) \, {\cal H}_{\perp}^{(g)\, [n];\, (2)}(1) + 4\, (1 - y)\, {\cal H}_{\varpar}^{(g)\, [n];\, (2)}(1) \right] \langle {\cal O}[n] \rangle \nt \\ & \phantom{=} \times 
    \int_{0}^{1} {\d \hat x'} \int_0^1 {\d \hat z}\, \left( \frac{1 - \hat z}{1 - \hat x^\prime} \right)^2\delta\big( {G(\hat x', \hat z)} \big)  \, \delta(1 - z) \nt \\
    & = \frac{\alpha}{4 y Q^2}  f^g (x; \mu^2)\sum_n \left[ \Big(1 + (1 - y)^2 \Big) \, {\cal H}_{\perp}^{(g)\, [n];\, (2)}(1) + 4\, (1 - y)\, {\cal H}_{\varpar}^{(g)\, [n];\, (2)}(1) \right] \langle {\cal O}[n] \rangle \nt \\ & \phantom{=} \times Q^2
    \int_{0}^{1} {\d \hat x'} \int_0^1 {\d \hat z}\, 
        \frac{(1 - \hat z)^2\big[ {(1 - \hat z)\, (M_\psi^2 + \hat z\, Q^2) + \hat z^2\, \qT^2} \big]}{\big[{(1 - \hat z)^2 M_\psi^2 + \hat z^2 \qT^2}\big]^2} {\delta (\hat x' - \hat x'_0)}  \, \delta(1 - z) \nt \\
    & \approx \frac{\alpha}{4 y Q^2}  f^g (x; \mu^2)\sum_n \left[ \Big(1 + (1 - y)^2 \Big) \, {\cal H}_{\perp}^{(g)\, [n];\, (2)}(1) + 4\, (1 - y)\, {\cal H}_{\varpar}^{(g)\, [n];\, (2)}(1) \right] \langle {\cal O}[n] \rangle \nt \\ & \phantom{=} \times Q^{2}
    \int_{0}^{1} {\d \hat x'} \int_0^1 {\d \hat z}\, 
        \frac{(1 - \hat z)^3}{\big[ {(1 - \hat z)^2 M_\psi^2 + \hat z^2 \qT^2} \big]^2} \left( M_\psi^2 + Q^2 \right) {\delta (\hat x' - \hat x'_0)}  \, \delta(1 - z) \nt \\
    & = \frac{\alpha}{4 y Q^2}
    f^g (x; \mu^2)\sum_n \left[ \Big(1 + (1 - y)^2 \Big) \, {\cal H}_{\perp}^{(g)\, [n];\, (2)}(1) + 4\, (1 - y)\, {\cal H}_{\varpar}^{(g)\, [n];\, (2)}(1) \right] \langle {\cal O}[n] \rangle
    \nt \\ & \phantom{=} \times
    \left[ \frac{ M_\psi^2 + Q^2}{M_\psi^2}\frac{Q^2}{2 M_\psi^2}\Big(\log \frac{M_\psi^2}{\qT^2} - 1 \Big) \right] \delta(1 - z) \nt \\
    & = \frac{\alpha}{4 y Q^2}  f^g (x; \mu^2)\sum_n \left[ \Big(1 + (1 - y)^2 \Big) \, {\cal H}_{\perp}^{(g)\, [n];\, (0)}(1,1) + 4\, (1 - y)\, {\cal H}_{\varpar}^{(g)\, [n];\, (0)}(1,1) \right] \langle {\cal O}[n] \rangle
    \nt \\ & \phantom{=} \times
    \left[ \frac{\hat x_{\rm max}}{2} \Big(\log \frac{M_\psi^2}{\qT^2} - 1 \Big) \right]  \, \delta(1 - z) \ , 
\label{eq: dsigma_C derivation}
\end{align}
{where we recall that $x = \xB/{\hat x_{\rm max}}$.}
{Note how the last lines in Eqs.~\eqref{eq: dsigma_B derivation} and~\eqref{eq: dsigma_C derivation} are respectively equivalent to what is presented in Eqs.~\eqref{eq: sigma B approximation} and~\eqref{eq: sigma C approximation}.}

\section{Eikonal method}
\label{app: eikonal method}
In this appendix we describe how to evaluate the soft gluon radiation from the leading order partonic subprocess in Eq.~\eqref{eq: partonic interaction LTM} by adopting the eikonal approximation.

\begin{figure}[t]
    \centering
    \includegraphics[width= 1.\linewidth, keepaspectratio]{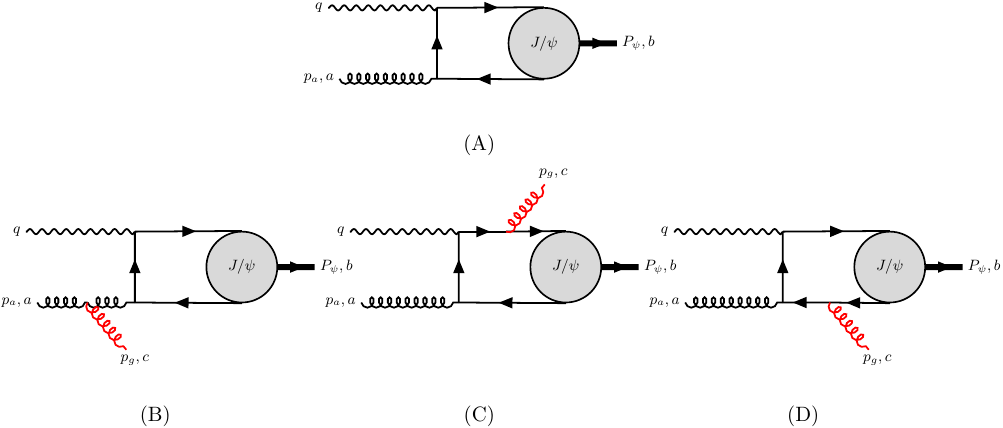}
    \caption{Leading order diagrams for the process $\gamma^*+g \to J/\psi(P_\psi)$, with the inclusion of soft gluon emission (in red) from initial and final states.}
    \label{fig: soft emission}
\end{figure}

The Born amplitude is depicted in Fig.~\ref{fig: soft emission}.A and the soft gluon emission is obtained by attaching a (soft) gluon to the initial (hard) gluon, as in Fig.~\ref{fig: soft emission}.B, or to the heavy quark-antiquark pair, Figs.~\ref{fig: soft emission}.C and~\ref{fig: soft emission}.D.
The eikonal gluon has a four-momentum $p_g$ that is negligible compared to the other (hard) momenta in the process. Hence, its polarization vector $\varepsilon_{\lambda_g}(p_g)$ fulfills the following relation
\begin{equation}
    \sum_{\lambda_g} \varepsilon^{*\alpha}_{\lambda_g}(p_g) \, \varepsilon^{\beta}_{\lambda_g}(p_g) \to - g^{\alpha\beta}\ .
\label{eq: completeness soft gluon}
\end{equation}
Moreover, the soft external gluon has color index $c$ and the initial gluon and the outgoing pair have color index $a$ and $b$, respectively.

The leading order amplitude of Fig.~\ref{fig: soft emission}.A is given by
\begin{equation}
    {\cal M}_0 = \delta_{ab}\, M_0\, ,
\end{equation}
with
{
\begin{equation}
    M_0 = \int \frac{\d^4 k}{(2\pi)^4}\,  \epsilon^{\mu}_{\lambda_a}(p_a)\, g_{\mu\nu}\, O^{\nu}_{ij}(P_\psi, k)\, \Phi_{ji}^{(b)}(P_\psi,k) \, ,
\label{eq: M0 NRQCD}
\end{equation}
 where $\epsilon_{\lambda_a}(p_a)$ is the polarization vector of the incoming gluon,
 $O(P_\psi, k)$ the perturbative operator related to the hard amplitude and $\Phi^{(b)}(P_\psi,k)$ the wave function of the non-relativistic $c\bar c$ pair. 
 Note that we are considering the pair having total momentum $P_\psi$ and relative momentum $2k$, and $i$ and $j$ are the color indices of the quark and antiquark, respectively.
 
The amplitudes for Figs.~\ref{fig: soft emission}.B-\ref{fig: soft emission}.D, where we have the insertion of an eikonal gluon, are identified by ${\cal M}_{1}^{(I)}$ with $I = B,\, C,\, D$.
They can be obtained from the Born one in Fig.~\ref{fig: soft emission}.A through proper replacements in a light-cone gauge. Here we list those needed to evaluate ${\cal M}_{1}$:

\begin{itemize}
\item {Emission from the incoming gluon}
\\
\begin{minipage}{.4\textwidth}
    \centering
    \includegraphics[width = .525\linewidth, keepaspectratio]{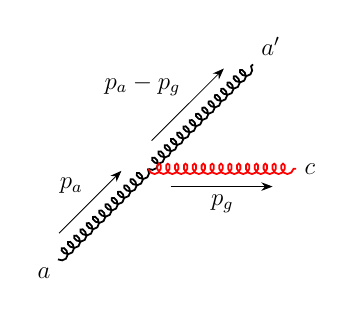}
\end{minipage}
\hfill
\begin{minipage}[t]{.55\textwidth}
    \raggedright
    $ \epsilon^\mu(p_a) \to - i\, g_s\, f_{aa'c}\, \epsilon^\mu(p_a) \left[ \big( p_a \cdot \epsilon^*_{\lambda_g}\big) / \big(p_a \cdot p_g \big) \right]$
\end{minipage}

\item {Emission from the outgoing quarkonia (solely color octet)}
\\
\begin{minipage}{.4\textwidth}
    \centering
    \includegraphics[width = .925\linewidth, keepaspectratio]{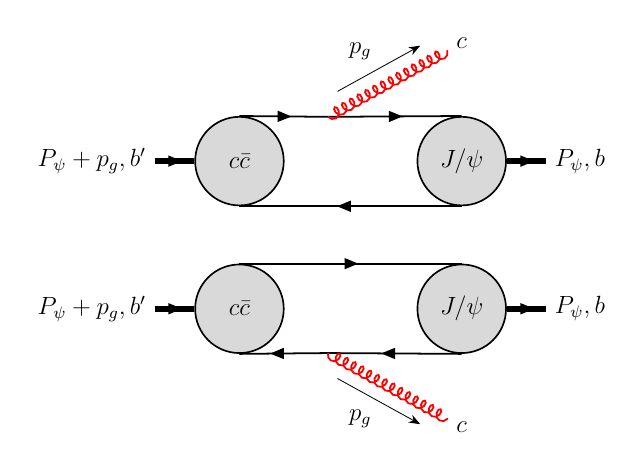}
\end{minipage}
\hfill
\begin{minipage}[t]{.55\textwidth}
$\Phi_{ji}(P_\psi, k) \to - i\, g_s\, f_{bb'c}\, \left[ \big( P_\psi \cdot \epsilon^*_{\lambda_g}\big) / \big(P_\psi \cdot p_g \big) \right] \Phi_{ji}(P_\psi, k)$
\end{minipage}
\end{itemize}

\noindent
From the first replacement we have that
\begin{align}
    {\cal M}_{1}^{(B)} & = \delta_{a'b}\,  M_{1}^{(B)} = \left( -i g_s\, f_{abc} \right) \varepsilon_{\lambda_g}^{*\alpha}(p_g) \left[ \frac{{p_a}_\alpha}{p_a \cdot p_g - i \epsilon}\right]  M_0\ ,
\label{eq: M1 B formula}
\end{align}
which is valid independently from the Fock-state of the $c \bar c$ pair, while from the second one we get
\begin{align}
    {\cal M}_{1}^{(C+D)} & = \delta_{ab'}\,  M_1^{(C+D)} = (i g_s\,  f_{abc}) \frac{{P_\psi}_\alpha}{P_\psi \cdot p_g}\,\varepsilon_{\lambda_g}^{*\alpha}(p_g)\, M_0\ ,
\label{eq: M1 C+D formula}
\end{align}
if the bound state is produced in a CO configuration (the relation is still independent of the other quantum numbers).
By combining Eqs.~\eqref{eq: M1 B formula} and~\eqref{eq: M1 C+D formula} we obtain the full amplitude that includes the soft gluon radiation from both the incoming gluon and the outgoing (CO)  $c \bar c$ pair, namely
\begin{equation}
    {\cal M}_1 = (i g_s\,  f_{abc}) \left[ \frac{{P_\psi}_\alpha}{P_\psi \cdot p_g} - \frac{{p_a}_\alpha}{p_a \cdot p_g} \right]\varepsilon_{\lambda_g}^{*\alpha}(p_g)\, M_0 \ .
\end{equation}
Averaging over colors and using Eq.~\eqref{eq: completeness soft gluon}, we then find that
\begin{equation}
    \overline{|{\cal M}_1|}^2 = g_s^2\, C_A \Big[ 2 S_g\big( p_a, P_\psi \big) - S_g\big( P_\psi, P_\psi \big) \Big]\, \overline{|{M}_0|}^2\, ,
\label{eq: M_1 amplitude squared}
\end{equation}
where
\begin{equation}
    S_g (v_1, v_2) = \frac{v_1 \cdot v_2}{\big( v_1 \cdot p_g\big) \big( v_2 \cdot p_g \big)}\ .
\end{equation}
 }
 
Considering a frame where $\bm q$ and $\bm p_a$ are along the $z$ axis, we can choose two light-cone vectors $\kappa_+^\mu$ and $\kappa_-^\mu$ such that
\begin{equation}
\begin{aligned}
    p_a^\mu & = \frac{\xB}{\hat x_{\rm max}}\kappa_+^\mu , \\
    q^\mu & = - \xB \kappa_+^\mu + \frac{Q^2}{2\, \xB} \kappa_-^\mu \ , \\
    p_g^\mu & = p_g^+ \kappa_+^\mu + p_g^- \kappa_-^\mu + p_{g \perp}^\mu = p_g^+ \kappa_+^\mu + p_g^- \kappa_-^\mu - P_{\psi \perp}^\mu\ ,
\end{aligned}
\end{equation}
where from the momentum conservation we have that $\bm P_{\psi\perp} = - \bm p_{g\perp}$, while by considering the softness of the gluon in the final state $(q + p_a)^2 \approx M_\psi^2$. 

We can introduce the variable $x_g$ defined by
\begin{align}
    x_g = \frac{p_g \cdot p_a}{q \cdot p_a}
        =  \frac{2\,\xB}{Q^2} p_g^- \ ,
\end{align}
which is also constrained by the momentum conservation
\begin{equation}
    \frac{\bm p_{g \perp}^2}{M_\psi^2 + Q^2} \leq x_g \leq 1 \ .
\label{eq: x_g range}
\end{equation}
Then, the phase space of the emitted (on-shell) soft gluon is given by
\begin{equation}
    \frac{\d^4 p_g}{(2\pi)^3} \delta\big(p_g^2\big) = \frac{\d^2 \bm p_{g\perp}}{2(2\pi)^3}\, \frac{\d p_g^-}{p_g^-} = \frac{\d^2 \bm P_{\psi\perp}}{2(2\pi)^3}\, \frac{\d x_g}{x_g}\ ,
\end{equation}
and the differential cross section will be proportional to the integration w.r.t $x_g$ of Eq.~\eqref{eq: M_1 amplitude squared}, namely\footnote{
The proportionality is due to the presence of Lorentz-invariant phase spaces, not explicitly shown here.
}
\begin{equation}
    \d \sigma_1 \propto \frac{g_s^2}{2(2\pi)^3}\, C_A \int_{\frac{\bm p_{g\perp}^2}{M_\psi^2 + Q^2}}^1 \frac{\d x_g}{x_g}\, \Big[ 2 S_g\big( p_a, P_\psi \big) - S_g\big( P_\psi, P_\psi \big) \Big]\, \overline{|{M}_0|}^2 = \frac{g_s^2}{2(2\pi)^3}\, C_A\, \Big[ 2\, I_a - I_\psi \Big]\, \overline{|{M}_0|}^2 \, .
\label{eq: integral soft gluon emission}
\end{equation}
The argument of the integral reads
\begin{align}
    S_g \big( p_a, P_\psi \big) & = \frac{p_a \cdot P_\psi}{\big( p_a \cdot p_g\big) \big( P_\psi \cdot p_g \big)}  \nt \\
    & \approx \frac{p_a \cdot q}{\big( p_a \cdot p_g\big) \Big[ \big( p_a \cdot p_g \big) + \big( q \cdot p_g \big) \Big]} = \frac{2}{M_\psi^2\, x_g^2 + \bm p_{g\perp}^2}\ ,
\end{align}
while
\begin{align}
    S_g \big( P_\psi, P_\psi \big) & = \frac{M_\psi^2}{\big( P_\psi \cdot p_g \big)^2} \approx \frac{M_\psi^2}{\Big[ \big( p_a \cdot p_g \big) + \big( q \cdot p_g \big) \Big]^2} = \frac{4\, M_\psi^2\, x_g^2}{\big( M_\psi^2\, x_g^2 + \bm p_{g\perp}^2 \big)^2}\ .
\end{align}
Hence, we can solve Eq.~\eqref{eq: integral soft gluon emission} analytically finding 
\begin{equation}
    I_a = \int_{\frac{\bm p_{g\perp}^2}{M_\psi^2 + Q^2}}^1 \frac{\d x_g}{x_g}\, \frac{2}{M_\psi^2\, x_g^2 + \bm p_{g\perp}^2} \approx \frac{1}{\bm p_{g\perp}^2} \left[\log \frac{M_\psi^2 + Q^2}{\bm p_{g\perp}^2} + \log \frac{M_\psi^2 + Q^2}{M_\psi^2} \right]
\end{equation}
and
\begin{equation}
    I_\psi = \int_{\frac{\bm p_{g\perp}^2}{M_\psi^2 + Q^2}}^1 \frac{\d x_g}{x_g}\, \frac{4\, M_\psi^2\, x_g^2}{\big( M_\psi^2\, x_g^2 + \bm p_{g\perp}^2 \big)^2} \approx \frac{2}{\bm p_{g\perp}^2}\ , 
\end{equation}
so that
\begin{equation}
    \d \sigma_1 \propto \frac{\alpha_s}{2\pi^2 \bm p_{g\perp}^2}\, C_A \left[\log \frac{M_\psi^2 + Q^2}{\bm p_{g\perp}^2} +\log \frac{M_\psi^2 + Q^2}{M_\psi^2} - 1 \right] \overline{|{M}_0|}^2 
\end{equation}
is in agreement with the first term of Eq.~\eqref{eq: logarithmic function}.

\bibliography{Bibliography}
\bibliographystyle{utphys}

\end{document}